# StarFinder: an IDL GUI based code to analyze crowded fields with isoplanatic correcting PSF fitting

E.Diolaiti[a], O.Bendinelli[a], D.Bonaccini[b], L.M.Close[b], D.G.Currie[b] and G.Parmeggiani[c]

[a] Università degli Studi di Bologna, Dipartimento di Astronomia
Via Ranzani 1, 40127 Bologna Italy

[b] European Southern Observatory
Karl-Schwarzschild Str. 2 - D-85748 Garching Germany

[c] Osservatorio Astronomico di Bologna
Via Ranzani 1, 40127 Bologna

**ABSTRACT**

StarFinder is a new code for the deep analysis of stellar fields, designed for well-sampled images with high and low Strehl ratio. It is organized in the form of a self-contained IDL widget-based application, with a 'user-friendly' graphic interface. We give here a general description of the code along with some applications to real data with space-invariant Point Spread Function (PSF). We present also some methods to handle anisoplanatic effects in wide-field Adaptive Optics (AO) observations.

## 1. INTRODUCTION

A stellar field can be represented as a superposition of stellar images on a smooth background emission, due to faint undetected stars, diffuse objects, etc. The image of a single point source can be either constant or spatially variable across the field.
The task of analyzing a crowded stellar field implies a reliable detection of the point sources and a determination of their position and flux. In the case of AO observations, the complex features of the PSF may induce false detections and affect the photometry of faint objects[1]. On the other hand an AO PSF is well sampled, i.e. the pixel dimension is not larger than ½ Full Width at Half Maximum (FWHM). The approach followed in StarFinder[2-4] exploits as much as possible the available knowledge of the PSF structure.
A major drawback of wide-field imaging with current AO systems is the presence of strong anisoplanatic effects[5]: off-axis stars appear blurred and radially elongated toward the reference source, commonly referred to as guide star. The analysis of such a stellar field requires the knowledge of the PSF in isoplanatic sub-patches of the imaged region. A general method to reconstruct the local PSF on the basis of recorded wavefront sensor data has been recently described[6].
In the present paper we describe the general features of the StarFinder code, along with some applications to real AO data with high and low Strehl ratio, characterized by a nearly spatially invariant PSF. Moreover we present our approach to the problem of anisoplanatism in AO images.
The code versatility has been also proved by an application to some NICMOS HST observations[7], which were analyzed by StarFinder and DAOPHOT[8].

Other author information: (Send correspondence to E.D.)
E. D.; Email: diolaiti@bo.astro.it; Telephone: +39 051 2095752; Fax: +39 051 2095700; Partly supported by the Italian Ministry for University and Research (MURST) under grant Cofin 98-02-32. O. B.; Email: bendinelli@bo.astro.it.
D. B.; Email: dbonacci@eso.org. L. M. C.; Email: lclose@eso.org. D. G. C.; Email: dcurrie@eso.org.
G. P.; Email: parmeggiani@bo.astro.it

## 2. CODE DESCRIPTION

### 2.1. Analysis procedure

The PSF is treated as a template for all the stars in the isoplanatic field, so its knowledge is fundamental for a reliable analysis. The determination of the PSF in StarFinder is obtained as median of a set of 'bona fide' uncontaminated star images, background-subtracted, centered with sub-pixel accuracy and normalized. The halo of the retrieved PSF is then smoothed, applying a median filtering technique with variable box size.

The analysis of a stellar field then is carried out forming a list of suspected stars selected as statistically significant intensity peaks above the background, which is estimated by interpolating a set of sky measurements relative to sub-regions arranged in a regular grid[9].

The presumed stars are listed by decreasing intensity and analyzed one by one. In order to illustrate a generic step of the algorithm, we assume that the first $n$ objects have already been examined and that a suitably scaled and positioned replica of the PSF has been put in a 'synthetic stellar field' for each star detected up to this point. The *(n+1)-th* object in the list might be a secondary PSF feature of a brighter star and this can be assessed by subtracting the local contribution of the brighter sources, known from the synthetic image. If a statistically significant residual can still be identified above the local background, it is compared to the PSF with a correlation check, as an objective measure of similarity. The object in hand is then accepted as a star if the correlation coefficient is greater than a selected level, and its position and flux are obtained by means of a local fit, performed on a small sub-image of size comparable to the diameter of the first diffraction ring of the PSF. The fitting model includes the contribution of the bright stars outside the fitting region known from the synthetic field, a slanting plane representing the local background and a sum of shifted weighted replicas of the local PSF, one for each point-source individuated in the fitting region. Sub-pixel astrometric accuracy is achieved by a non-linear optimization of the star positions, which is based on the interpolation of the given PSF array. A similar technique has been described by Véran and Rigaut[10]. If the fit is acceptable the star field catalogue and the related syntethic image are upgraded by the new entry.

The sequence of steps described above (re-identification by subtraction, correlation check, local fitting, synthetic field upgrading) is performed for each object in the initial list. At the end of the analysis one may perform a new search for previously lost objects (e.g. secondary components of close binary stars, faint stars detectable only at a lower correlation level, etc.), removing the contribution from all the stars detected up to this point. It should be stressed that this is just a tool to highlight significant residuals, which are clues of an incorrect detection or of the need to take into account a new object to be analyzed. Any further procedure operation, performed on the original frame, is aimed to eliminate the effects arising from the wings of near luminous stars or other possible sources.

### 2.2. IDL code and user interface

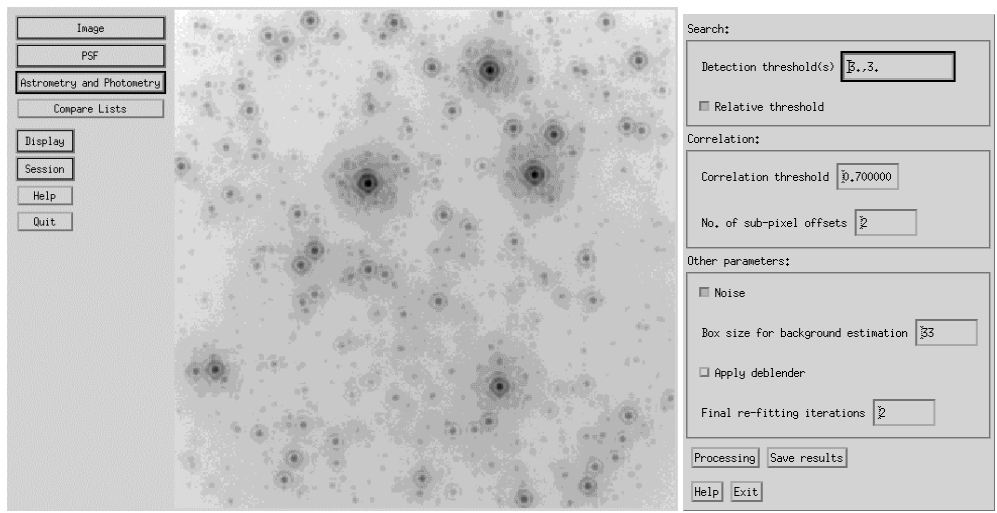

**Figure 1.** Left: Main widget. Right: widget form to define the parameters for stellar fields analysis.

The StarFinder algorithm was originally created as an IDL program to analyze a single stellar field image. Then it has been provided with a collection of auxiliary routines for data visualization, PSF extraction, noise estimation, basic image processing, in order to allow the user to analyze a stellar field, produce an output list of objects and compare different lists, e.g. referred to different observations of the same target.

A widget-based graphical user interface has been created. The main widget appearing on the screen (Fig.1) is an interface to other widget-based applications, each corresponding to a particular operation (e.g. PSF estimation, stars detection, comparison of lists, etc.). The secondary applications block the access to the main widget, in order to avoid 'cross-talks', which might corrupt the data associated to the current session. The basic documentation about the code can be found in the on-line help pages.

The code is entirely written in the IDL language and has been tested on Windows and Unix platforms supporting IDL v. 5.0 or later.

## 3. APPLICATIONS WITH SPACE-INVARIANT PSF

### 3.1. High Strehl case

The algorithm has been run on a K-band PUEO image of the Galactic Center (Fig. 2), kindly provided by François Rigaut. This image is an example of a well-sampled high-Strehl AO observation of a stellar field.

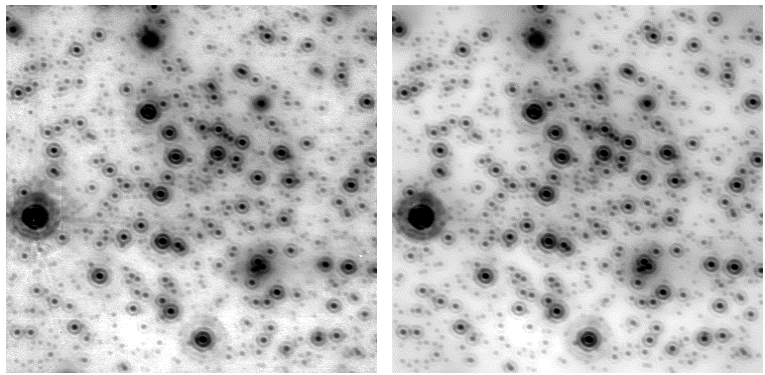

**Figure 2.** Left: PUEO image of the Galactic Center. Right: reconstructed image, given by the sum of ~ 1000 detected stars and the estimated background. The display stretch is logarithmic.

The total field of view is about 13×13 $arcsec^2$ and the pixel size 0.035 *arcsec*. The PSF is nearly stable across the image, apart from some minor features especially on the shape of the first diffraction ring. We will show that assuming a constant PSF it is possible to obtain accurate results.

About 1000 stars have been detected, with a correlation coefficient of at least 0.7; the reconstructed image is shown in Fig. 2. We have evaluated the accuracy of the algorithm by means of an experiment based on the addition of artificial stars: for each magnitude bin in the retrieved luminosity function (Fig. 3), 10% of synthetic stars at random positions have been added to the original image. In practice 10 frames have been created with this procedure and analyzed separately. The catalogues of detected artificial stars for each frame have been merged together; then the astrometric and photometric errors have been computed and plotted as a function of the true magnitude. The plots show accurate astrometry and photometry and there is no apparent photometric bias. It is interesting to consider the accuracy for the stars brighter than a relative magnitude of 5: the mean astrometric error is <0.5 *mas* and the mean absolute photometric error is <0.01 *mag*. The artificial sources are contaminated by the background noise already present in the observed data and by the photon noise due to neighboring stars; no additional photon noise has been added, so the error estimates should be regarded as lower limits.

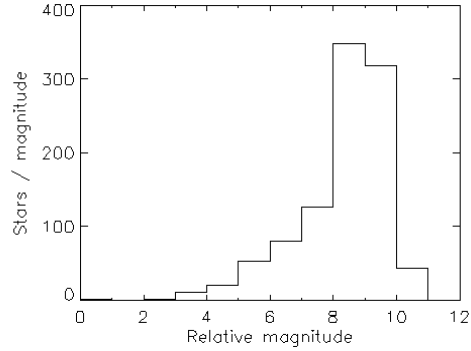

**Figure 3.** Estimated luminosity function.

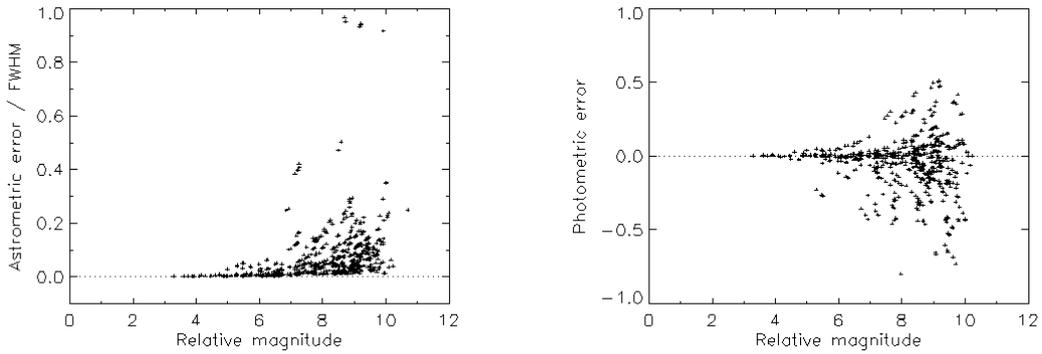

**Figure 4.** Left: plot of astrometric errors vs. relative magnitude of detected synthetic stars; the errors are quoted in FWHM units (1 FWHM ≅ 4 pixel) and represent the distance between the true and calculated positions. Right: plot of photometric errors. A tolerance of 1 PSF FWHM has been used to match each detected star with its true counterpart.

### 3.2. Low Strehl case

A low-Strehl example is represented by a K-band image of the globular cluster 47Tuc, observed at the ESO 3.6m telescope with the ADONIS AO system (Fig. 5). The PSF is very well sampled, with a FWHM of ~ 6 *pixels* (1 *pixel* = 0.1 *arcsec*).
As in the previous high-Strehl case, we have evaluated the accuracy of the algorithm by means of an experiment with synthetic stars. The astrometric and photometric errors of the detected artificial sources are shown in Fig. 6. About 40% of the detected synthetic stars have an astrometric error smaller than 0.1 PSF FWHM and photometric accuracy better than 0.1 *mag*; in the previous example the corresponding percentage is 56%: we attribute the lower accuracy in the low-Strehl case to the worse image quality.

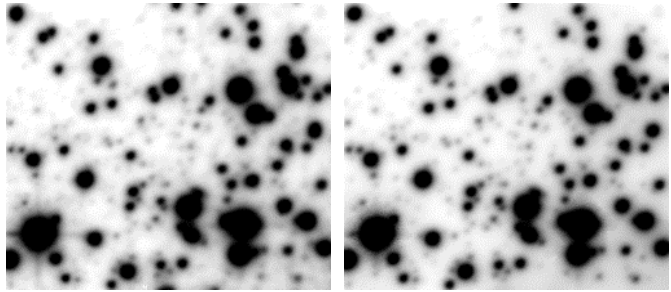

**Figure 5.** Left: 47Tuc observed image. Right: reconstruction. The display stretch is logarithmic.

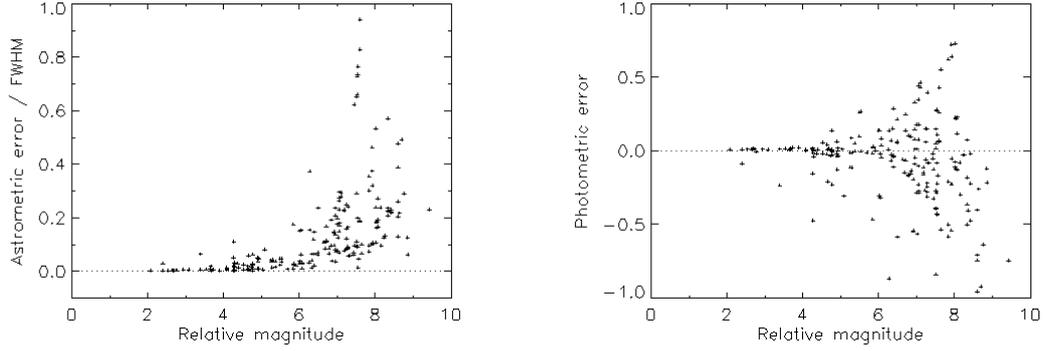

**Figure 6.** Left: plot of astrometric errors vs. relative magnitude of detected synthetic stars; the errors are quoted in FWHM units (1 FWHM ≅ 6 pixel) and represent the distance between the true and the calculated position. Right: plot of photometric errors. A tolerance of 1 PSF FWHM has been used to match each detected star with its true counterpart.

## 4. ANISOPLANATIC EFFECTS

### 4.1. Statement of the problem

An example of a wide-field AO observation is shown in Fig. 7: it represents a 1′×1′ image of the Trapezium cluster, obtained with the CFHT. A comparison between the guide star and the off-axis star shown in the same figure clearly shows the strong anisoplanatic effect affecting this image: the Strehl-ratio decreases by a factor of ~3 and the FWHM increases by a factor of ~2.

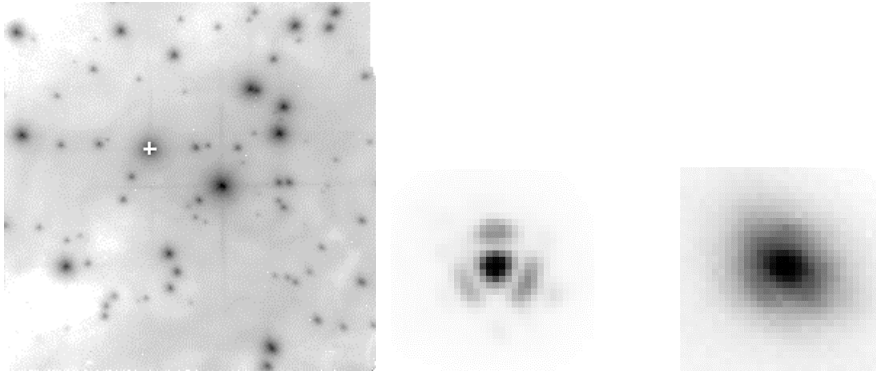

**Figure 7.** Left: Trapezium cluster, observed in the K-band with the CFHT; the cross indicates the guide star. Center: guide star. Right: off-axis star.

The relatively small angular size of the region of sky corrected by current AO systems is well known. A significant improvement in this sense will be possible with multi-conjugate AO systems[11] and 3D atmospheric tomography techniques[12]. The main result of a theoretical analysis[13, 6] of the anisoplanatic effect in current AO systems is that the off-axis PSF can be expressed as the convolution of the on-axis PSF (guide star) with a spatially variable kernel, i.e.

$$PSF(\mathbf{r},\alpha) = PSF(\mathbf{r},0) * K(\mathbf{r},\alpha). \tag{1}$$

In the following three subsections we describe our approaches to the problem of the PSF spatial variation, based on equation 1.

## 4.2. Modeling the PSF variation

The first method is based on a parametric modeling of the anisoplanatic kernel: this would be an appealing solution, because such a model could be calibrated using a few reference point sources and could then be used to reproduce the PSF at any location in the field of view.

The simplest model can be obtained considering the effect of residual tip-tilt, arising from the partial correction of the instantaneous image-motion. A long exposure PSF degraded by uncorrected instantaneous image-motion can be represented as the convolution of the short-exposure PSF with a bivariate gaussian[14], whose standard deviation is related to the RMS of the instantaneous jitter. On the other hand the major effect observed in the Trapezium image is a radial elongation of the off-axis PSF. This suggests to represent the anisoplanatic kernel associated to the residual tip-tilt with a simple elliptical gaussian elongated toward the guide star.

We have applied this method to the Trapezium image, fitting many stars at various radial distances with a model given by the convolution of the guide star with a suitable elliptical gaussian. The measurements of the radial and lateral standard deviation of the kernel are plotted in Fig. 8, as a function of the distance from the reference source. A linear fit to the measurements, excluding the constant term of the polynomials, yields a ratio of the radial to the lateral elongation of about 1.81, 5% larger than the theoretical value[15] of $\sqrt{3}$. This is a good agreement. Nevertheless the measurements show a considerable scatter. Moreover the measured values of radial elongation near the origin (see the first plot in Fig. 8) are systematically below the best fit line. Finally we have found that the estimated values of the parameters tend to increase with the size of the sub-image used for the estimation: this suggests that a single component model, represented by an elliptical gaussian, cannot represent the kernel at different radial distances from its center. A direct deconvolution of an off-axis observed PSF from the guide star produces a kernel with a central peak superposed on a wider irregular 'halo'. Probably a two-component model (e.g. the superposition of two scaled gaussian functions of different width) would represent a substantial improvement. We leave this point to further investigation.

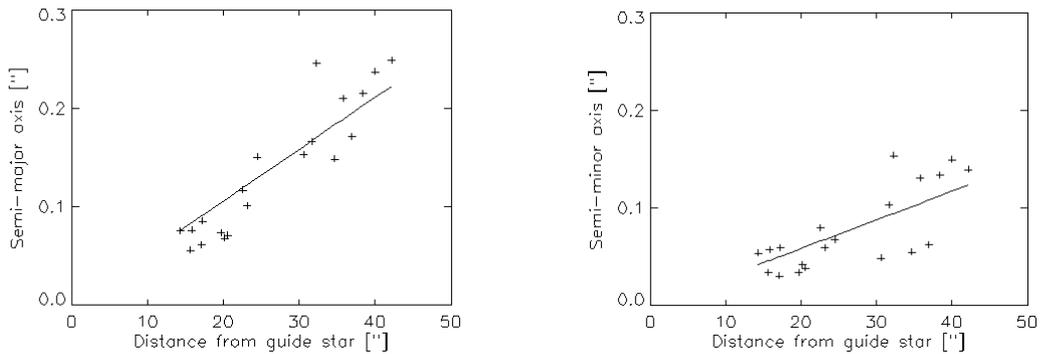

**Figure 8.** Left: radial elongation of convolving gaussian. Right: transversal width.

## 4.3. Photometry with a set of local PSF estimates

Another approach is the partition of the stellar field into a regular grid of nearly isoplanatic sub-regions and the PSF determination for each patch: this corresponds to a piece-wise constant approximation of the anisoplanatic kernel. StarFinder is then applied to analyze suspected stars using the appropriate PSF in each field partition.

In order to test the accuracy achievable with this approach, we have simulated a stellar field with 1000 point sources randomly placed, extracting the relative magnitudes from an observed luminosity function with a dynamic range of 8 *mag*. The point sources have been convolved with a high-Strehl critically sampled synthetic PSF of the ESO ADONIS system. The simulated image with space-invariant PSF is shown in Fig. 9. Then we have generated a similar field with space-variant PSF, given by the convolution of the reference PSF with an elliptical gaussian elongated toward a reference position, corresponding to the bright star at the center of the image in Fig. 9. The coefficients (radial and transversal standard deviation) of the anisoplanatic kernel depend linearly on the distance from the guide star and have a constant ratio of $\sqrt{3}$; the slope of the linear polynomials expressing these coefficients as a function of the radial distance has been chosen in order to have a Strehl ratio reduction between the guide star and a source in the corner of the image of a factor ~2.4 and a corresponding increase in the mean PSF FWHM of a factor ~1.9 (see axial plots in Fig. 9).

The stellar fields have been contaminated with photon and gaussian noise: the signal to noise ratio of the fainter sources, expressed by the ratio of the peak intensity of the star to the overall noise standard deviation in the same pixel, is not larger than 10.

The goal of the present analysis is to compare the photometric accuracy obtainable with the same stellar population in the isoplanatic and anisoplanatic case, assuming here known the local PSF on a grid of $N{\times}N$ sub-regions. The results are shown in Table 1 and Fig. 10. The accuracy in the space-variant case increases with the number of local measurements of the PSF: we expect that in the limit of a very dense partitioning of the imaged field, it would tend to the one achievable in the space-invariant case. In general the local PSF has to be extracted directly from the image, as a superposition of neighboring stars; we are developing an automatic procedure for this purpose.

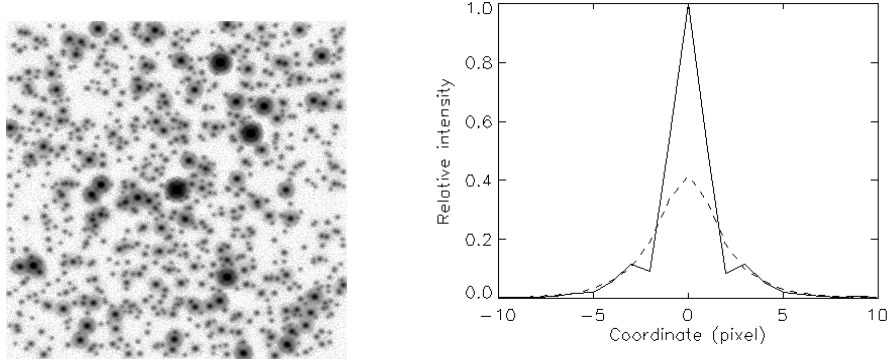

**Figure 9.** Left: simulated stellar field with space-invariant PSF. The image size is 512×512. Right: axial plot of on-axis (continuous line) and off-axis (dashed line) PSF in the corresponding simulated field with space-variant PSF.

| Case | | % detected | % false | |Δmag| |
|---|---|---|---|---|
| S-I | | 98 | 0.0 | 0.030 |
| S-V | 3×3 | 95 | 0.2 | 0.089 |
| S-V | 5×5 | 96 | 0.5 | 0.066 |
| S-V | 7×7 | 96 | 0.2 | 0.056 |

**Table 1.** Detection reliability and photometric accuracy for the space-invariant (S-I) and the space-variant (S-V) case, assuming to know the local PSF on a regular grid $N{\times}N$, for $N$ = 3, 5, 7.

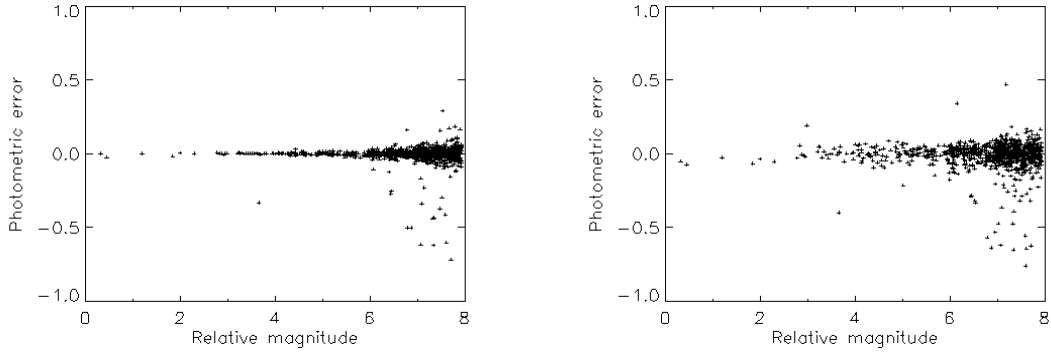

**Figure 10.** Left: photometric accuracy on simulated field with space-invariant PSF. Right: photometric accuracy on simulated field with space-variant PSF, measured on a 7×7 grid.

**4.4. Removing the anisoplanatic effect by space-variant deconvolution**

The relation between the off-axis and the on-axis PSFs (Eq. 1) is the basis of a method to remove the anisoplanatic effect from an AO observation.

Given an intensity distribution $x(\alpha',\beta')$ representing the target object, namely a sum of weighted delta-functions in the case of a stellar field, the image formed by a linear system is given by the well-known Fredholm superposition integral

$$y(\alpha,\beta) = \iint h(\alpha,\beta,\alpha',\beta')\, x(\alpha',\beta')\, d\alpha' d\beta' , \qquad (2)$$

where $h(\alpha, \beta, \alpha', \beta')$ is the response in $(\alpha, \beta)$ to a point source placed in $(\alpha', \beta')$. A discretization of the above integral leads to the matrix equation

$$\mathbf{y} = \mathbf{H}\mathbf{x}, \qquad (3)$$

where the array $\mathbf{H}$ is block-circulant[16] in the space-invariant case.
According to equation 1 we can express $\mathbf{H}$ in the form

$$\mathbf{H} = \mathbf{K}\mathbf{H}', \qquad (4)$$

where $\mathbf{H}'$ is the block-circulant array corresponding the on-axis PSF and $\mathbf{K}$ represents the anisoplanatic kernel.
Replacing equation 4 in 3 we obtain

$$\mathbf{y} = (\mathbf{K}\mathbf{H}')\mathbf{x} = \mathbf{K}(\mathbf{H}'\mathbf{x}) = \mathbf{K}\mathbf{y}', \qquad (5)$$

where $\mathbf{y}' = \mathbf{H}'\mathbf{x}$ is the image as if it were obtained by an isoplanatic system.

In order to find the unknown intensity distribution $\mathbf{y}'$ we must be able to deconvolve the observed image from the space-variant kernel $\mathbf{K}$. Since the array $\mathbf{K}$ has no particular structure, we cannot invert it, but we have to rely on iterative deconvolution algorithms based only on the application of the direct blur. An example is the accelerated Richardson-Lucy (R-L) algorithm[17], given by the iterative scheme

$$\mathbf{y}'_{i+1} = \mathbf{y}'_i + \alpha_i \, \Delta\mathbf{y}'_i, \qquad \Delta\mathbf{y}'_i = \mathbf{y}'_i \left[ \mathbf{K}^T \left( \frac{\mathbf{y}}{\mathbf{K}\mathbf{y}'_i} \right) - 1 \right]. \qquad (6)$$

If the blurring kernel is known on a partition of the image domain, the application of the blurs $\mathbf{K}$ and $\mathbf{K}^T$ in the R-L method can be efficiently performed by means of the 'overlap & save' and 'overlap & add' schemes[18]. In practice the space-variant blur is applied to non-overlapping sub-regions of the object of interest, taking into account the 'interaction' effects between adjacent patches; in this way the object is treated as a whole, preventing the onset of 'sewing' effects. Piece-wise constant and linear interpolation schemes of the local measurements of the blurring function can be adopted. The 'overlap & save' and 'overlap & add' algorithms have been coded in IDL and inserted in the R-L method.

In the case of a stellar field, it is possible to determine the space-variant kernel $\mathbf{K}$ by partitioning the observed image into nearly isoplanatic sub-patches and then deconvolving the local PSF for each sub-region from the guide star. The set of measurements of the anisoplanatic kernel can then be fed into the R-L scheme for the space-variant de-blurring of the observed image.

The procedure described above has been applied to the simulated field described in Sect. 4.3, assuming to know the local PSF on a 7×7 partition of the image. A comparison of an axial profile across the guide star and an off-axis PSF before and after removing the anisoplanatic effect is shown in Fig.11. In order to assess the photometric reliability of the corrected image, we have applied the space-invariant StarFinder. The photometric analysis has been performed with a PSF estimate extracted directly from the corrected field, to average out the residual PSF variations which would not be accounted for using the guide star. The photometric errors for the detected stars are plotted in Fig. 11: the mean absolute error is ~0.058 *mag.*, comparable to the results obtained in Sect. 4.3 by means of the space-variant StarFinder, and the photometry appears to be substantially unbiased. The detection rate is the same (~98%), even though we report a slightly higher percentage of false detections (~1%), which might be a consequence of the noise propagation in the deconvolution process.

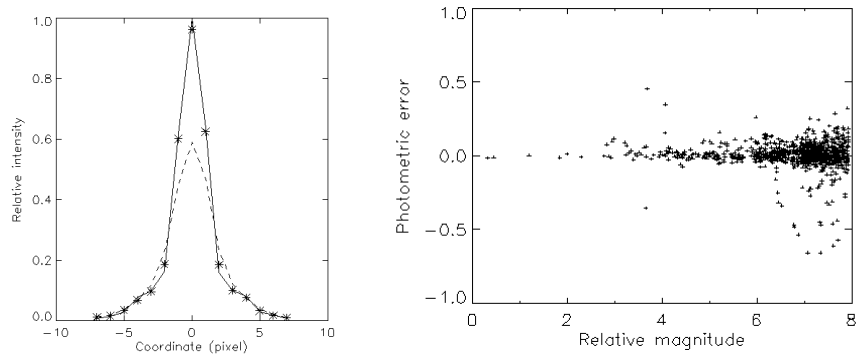

**Figure 11.** Left: axial plot of an off-axis star in the space-invariant image (continuous line), in the space-variant image (dashed line) and in the reconstructed image (star symbols). Right: photometric accuracy on the reconstructed image.

## 5. CONCLUSIONS

The StarFinder code seems to be able to analyze adequately sampled high- and low- Strehl images of crowded fields with high accuracy, and the working version up to this point is available under request to E.D. Its implementation aimed at the choice of a suitable strategy in analyzing images with space variant PSF is in progress, strongly encouraged by the preliminary results obtained on anisoplanatic images.

## ACKNOWLEDGMENTS

François Rigaut is acknowledged for kindly providing the PUEO image of the Galactic Center and for supporting the initial development of this method.

## REFERENCES

1. O. Esslinger, M. G. Edmunds, "Photometry with adaptive optics: A first guide to expected performance ", *Astron. & Astrophys. Suppl. Ser.,* **129**, pp.617-635, 1998.
2. E. Diolaiti, O. Bendinelli , D. Bonaccini, G. Parmeggiani and F. Rigaut , "An algorithm for crowded stellar fields analysis", in *Proceedings of the ESO/OSA meeting on Astronomy with Adaptive Optic*s, D. Bonaccini, ed., pp.175-184, ESO, Garching, 1998.
3. E. Diolaiti, O.Bendinelli, D. Bonaccini, L. M. Close, D.G. Currie and G.Parmeggiani, "StarFinder: a code for crowded stellar fields" in *Proceedings of ADASS IX,* D. Crabtree, N. Manset and C. Veillet , eds., ASP Conference Series, San Francisco, 1999. In press.
4. E. Diolaiti, O. Bendinelli, D. Bonaccini, L. M. Close, D.G. Currie and G.Parmeggiani, "StarFinder: a new code for stellar fields analysis". In preparation.
5. L. M. Close, F. Roddier, D. Potter, C. Roddier, J. E. Graves, M. Northcott, "Astronomy with adaptive optics: experiences from the University of Hawaii AO program", in *Proceedings of the ESO/OSA meeting on Astronomy with Adaptive Optic*s, D. Bonaccini, ed., pp.109-120, ESO, Garching, 1998.
6. T. Fusco, J. -M. Conan, L. M. Mugnier, V. Michau, G. Rousset, *Astron. & Astrophys. Suppl. Ser.,* **142**, pp. 149-156, 2000.
7. A. Aloisi, M. Clampin, E. Diolaiti, L. Greggio, C. Leitherer, A. Nota, L. Origlia, G. Parmeggiani and M. Tosi, "The red stellar population in NGC 1569". In preparation.
8. P. Stetson, "DAOPHOT - a computer program for crowded-field stellar photometry", *Pubbl. Astr. Soc. Pac.*, **99**, pp.191-222, 1987.
9. E. Bertin and S. Arnouts, "SExtractor: software for source extraction ", *Astron & Astrophys. Suppl. Ser.,* **117**, pp.393-404, 1996.
10. J.-P. Véran, F. Rigaut, "A deconvolution method for accurate astrometry and photometry on adaptive optics imagers of stellar fields", Proc. SPIE**, 3353**, pp. 426-437, 1998.
11. B. Ellerbroek, F. Rigaut, "Optics adapt to the whole sky", *Nature,* **403**, pp.25-26, 2000.
12. R. Ragazzoni, E. Marchetti and G. Valente, *Nature,* **403**, pp.54-56, 2000.

13. V. V. Voitsekhovich and S. Bara, "Effect of anisotropic imaging in off-axis adaptive astronomical systems", *Astron & Astrophys. Suppl. Ser.,* **137**, pp.385-389, 1999.
14. F. Roddier, "Optical propagation and image formation through the turbulent atmosphere " in *Diffraction-limited imaging with very large telescopes,* D.M.Alloin and J.-M. Mariotti, eds., pp.33-52, Kluwer Academic Publishers, Dordrecht, 1988.
15. R. J. Sasiela and J. D. Shelton, " Transverse spectral filtering and Mellin transform techniques applied to the affect of outer scale on tilt and tilt anisoplanatism", *J. Opt. Soc. Am. A* ,**10**, pp.646-660, 1993.
16. R. C. Gonzalez and R. E. Woods, *Digital image processing*, Addison-Wesley Publishing Company, Reading, 1992.
17. H. -M. Adorf, R. N. Hook, L. B. Lucy and F. D. Murtagh, "Accelerating the Richardson-Lucy restoration algorithm ", in *4$^{th}$ ESO/ST-ECF Data Analysis Workshop*, P. Grosbol and R. C. E. Ruijsscher, eds., pp.99-103, ESO, Garching,1992.
18. J. G. Nagy and D. P. O'Leary, "Fast iterative image restoration with a spatially-varying PSF", in *Advanced Signal Processing Algorithms, Architectures, and Implementations VIII*, F.T. Luk, ed., pp.388--399, SPIE, Bellingham, 1997.